\documentclass[twocolumn]{article}
\usepackage{pkuarticle}
\AtBeginDocument{%
  }

\usepackage{array}
\usepackage{algorithm}
\usepackage{algpseudocode}
\usepackage{amsmath}
\usepackage{amsfonts}
\usepackage{booktabs}
\usepackage{graphicx}
\usepackage{makecell}
\usepackage{tabularx}
\usepackage{enumitem}
\usepackage{textcomp}
\usepackage{url}
\usepackage[numbers]{natbib}

\definecolor{pktred}{HTML}{F57C6E}
\definecolor{schgreen}{HTML}{84C3B7}
\definecolor{resblue}{HTML}{71B7ED}
\definecolor{graydot}{gray}{0.6}

\newcommand{\circred}{\tikz\draw[fill=pktred,   draw=pktred]   (0,0) circle (2.5pt);}
\newcommand{\circgreen}{\tikz\draw[fill=schgreen, draw=schgreen] (0,0) circle (2.5pt);}
\newcommand{\circblue}{\tikz\draw[fill=resblue,  draw=resblue]  (0,0) circle (2.5pt);}
\newcommand{\circgray}{\tikz\draw[fill=graydot, draw=graydot] (0,0) circle (2.5pt);}

\newcommand{\threecir}[3]{#1\ #2\ #3}

\newcommand{\codename}{\texttt{C2C-Explorer}}

\renewcommand{\and}{,\ }
\title{C2C-Explorer: An Exploration Framework for Chip-to-Chip Interconnect Architectures in LLM Cloud Computing Systems}

\author{Jiayi Li$^{\triangle\dagger\diamond\square}$, Di Wu$^{\diamond\square}$, Qingxu Li$^{\star}$, Hongxiao Zhao$^{\triangle\dagger}$, Jiaqi Yang$^{\triangle\dagger}$, Anjunyi Fan$^{\triangle\dagger}$, Wenbin Zhang$^{\diamond\square}$, Boqiang Wu$^{\diamond\square}$, Shuting Liu$^{\diamond\square}$, Shifeng Fang$^{\diamond\square}$, Jianbo Dong$^{\star}$, Dimin Niu$^{\diamond\square}$ and Bonan Yan$^{\triangle\dagger\ast}$}
\affiliation{$^\triangle$Institute for Artificial Intelligence, Peking University, Beijing, China; $^\dagger$Beijing Advanced Innovation Center for Integrated Circuits, School of Integrated Circuits, Peking University, Beijing, China; $^\diamond$Hupan Lab, Hangzhou, China; $^\square$Damo Academy, Alibaba Group, Hangzhou, China; $^\star$Alibaba Cloud, Alibaba Group, Beijing, China\\
\small\texttt{bonanyan@pku.edu.cn}}

\newcommand{\copyrightnotice}{%
  \begingroup
  \renewcommand{\thefootnote}{}%
  \footnotetext{\footnotesize
    This paper has been accepted for publication at the 63rd ACM/IEEE
    Design Automation Conference (DAC '26), July 26--29, 2026, Long Beach,
    CA, USA.}%
  \endgroup
}

\begin{document}
\maketitle
\copyrightnotice

\begin{abstract}

The scaling-up of large language models (LLMs) necessitates computing systems to have multi-processor-chip architectures, elevating the importance of chip-to-chip (C2C) communication.
However, designing efficient C2C hardware architectures for LLM workloads faces three key challenges: generating realistic LLM-specific C2C traffic, accurately simulating hardware-level communication at scale, and efficiently exploring the exponentially large C2C design space.
We propose \codename{}, an adaptive Bayesian DSE framework that integrates a LLM-workload-driven traffic generator, a scalable interconnect simulator (switch/full-mesh, up to 512 chips), and a metric-guided evaluator into a workload-to-hardware optimization pipeline, enabling systematic C2C
architectural co-design under realistic LLM workloads. Validated against FPGA-based C2C prototypes, the C2C simulator achieves 2.46–8.23\% end-to-end timing error across diverse traffic patterns. Its hybrid cycle \& event model further accelerates large-scale simulation by up to 7.8$\times$ over a pure cycle-accurate baseline. Applied to a 32-XPU DeepSeek-R1-671B inference workload, \codename{} identifies configurations that improve goodput by 44.1\% and reduce memory by 98.4\%.
\codename{} is open-source and available at \url{https://github.com/Selinaee/C2C-Explorer}.

\end{abstract}

\noindent\textbf{Keywords:} chip-to-chip communication, supernode, scale out, LLM simulator, scale up, large language model, cloud computing system

\section{Introduction}

\begin{figure}[t]
\centering
\includegraphics[width=\linewidth]{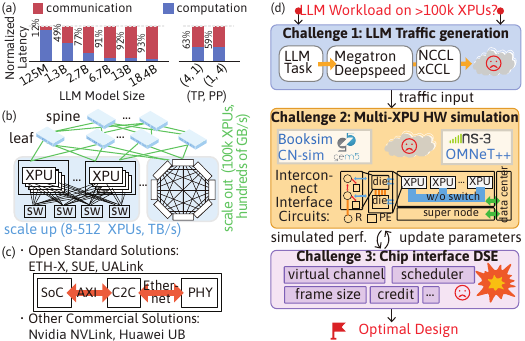}
\caption{
(a) Left: latency breakdown (communication \& compute) per training iteration vs. LLM model sizes~\cite{intto1}. Right: latency breakdown for Llama3.1-8B inference under various parallelism settings~\cite{xu2025characterizingcommunicationpatternsdistributed}.
(b) Typical interconnect architecture for multi-XPUs LLM cloud computing systems.
(c) Representative C2C interconnect protocols.
(d) Three challenges in C2C design for LLM computing systems.}
\label{fig: introduction}
\end{figure}

With the exponential growth in parameter size of large language models (LLMs), deployment strategies have transitioned from single-accelerator setups to coordinated multi-node, multi-accelerator systems that form super-nodes for enhanced resource coordination~\cite{Kim2019CoDesign2.5D}. This shift has revealed that inter-chip communication increasingly dominates computation in large-scale LLM workloads. Specifically, recent studies demonstrate that under typical parallel affinity configurations, communication accounts for over 90\% of iteration time during training (Figure~\ref{fig: introduction}(a), left)  and exceeds 50\% of total latency during inference (Figure~\ref{fig: introduction}(a), right), emphasizing the critical demand for high-bandwidth, low-latency interconnects~\cite{liao2025ub,MaoYLLC19,MudassarKM18}.

Figure~\ref{fig: introduction}(b) illustrates a typical interconnect architecture employed in LLM computing systems with multiple XPUs\footnote{XPUs refer to various architectures of processing units, e.g., central processing units (CPU), graphics processing units (GPU), neural processing units (NPU), tensor processing units (TPU), and FPGAs.}.


To achieve optimal deployment of LLM workloads on computing systems with $>10^5$ XPUs, it is essential to determine the optimal interconnect in the scale-up domain. However, this endeavor faces significant challenges of lacking appropriate EDA toolkits for:
\begin{enumerate}[leftmargin=*,label=(\alph*)]
\item \textbf{LLM-specific traffic generation}.
Figure~\ref{fig: introduction}(d) demonstrates that various LLM parallelization strategies, such as data or tensor parallelism, produce distinct collective communication patterns in frameworks like Megatron~\cite{megatron} and DeepSpeed~\cite{deepspeed}, with each configuration inducing specific operations (e.g., all-reduce, all-to-all) via NCCL~\cite{NCCL} to derive GPU-to-GPU C2C traffic traces.
To obtain realistic C2C flows, these traces must be decomposed according to the physical interconnect structure and port mapping, a process that is inherently complex and topology-dependent.

\item \textbf{Multi-XPU hardware simulation}.
Emulation of interconnect behavior for scaled-out systems with scaled-up chips is missing because of the complicated cross-layer protocols.

\item \textbf{Design space exploration for C2C interface}.
Various factors, such as LLM workloads, network topologies, and chip-to-chip (C2C) link configurations, interact with multiple submodels and hardware parameters, including virtual channels (VCs), schedulers, frame sizes, and credit sets, resulting in a combinatorial explosion.
Consequently, evaluating and co-optimizing the end-to-end performance across all possible combinations remains a formidable challenge.
\end{enumerate}

To fill this gap, this work presents \codename{}, a joint framework for interconnect hardware design space exploration and communication simulation under LLM workloads. The major contributions include:
\begin{itemize}[leftmargin=*]
\item \textbf{An LLM traffic generator} under practical LLM workloads that bridges P2P (e.g. GPU-to-GPU) communication to C2C systems by producing AXI-accurate SoC-to-C2C traffic input and feeding it directly into our C2C simulator. It integrates dual-layer flow-control semantics: a combination of application-level pacing and link-level credit backpressure, enabling realistic cycle-accurate C2C timing under real LLM workloads.

\item \textbf{An open-source, fast, and scalable interconnect simulator} based on AXI and Ethernet PHY, supporting switch and full-mesh topologies with up to hundreds of XPUs and thousands of C2C links. A hybrid modeling strategy—cycle-accurate C2C port modeling combined with event-driven switch scheduling—provides substantial speedup while preserving link-level fidelity.

\item \textbf{An adaptive Bayesian design space exploration (AB-DSE) method} that applies hardware-informed feasibility constraints (derived from our analysis of core C2C parameters) to prune the combinatorial space, and performs Bayesian optimization to efficiently identify the optimal hardware configuration~\cite{AlawiehWL17}~\cite{WangYYZH17}.

\end{itemize}

To validate \codename{}, we employ LLM and other communication workloads, using real-world FPGA-based C2C host prototypes as a baseline for comparison. The simulator achieves consistently low end-to-end timing error (2.46–8.23\%) across multiple traffic patterns. Moreover, the hybrid scheduling model accelerates simulation runtime by up to 7.8$\times$. With verified \codename{} toolkits, we systematically analyze three core C2C design parameters (packetization, scheduling, and resource-allocation) and evaluate them against standard performance metrics. This analysis reduces the valid configuration space from 2394 to 1152 and achieves convergence to the optimum in approximately 20 iterations under a 32-XPU LLM workload.
{\codename} is available at \url{https://github.com/Selinaee/C2C-Explorer}).


\section{Related Works}

On one hand, existing LLM-oriented simulation frameworks such as Vidur~\cite{agrawal2024vidur}, SimAI~\cite{simai}, and vTrain~\cite{vtrain} primarily focus on system-level performance evaluation of existing GPU infrastructures, providing valuable insights into large-scale deployment and scheduling behaviors.
Among them, SimAI can, given an LLM workload and parallelization strategy, simulate collective communication algorithms and NCCL primitives to produce GPU-to-GPU flow traces, but these traces do not map to physical hardware paths and thus cannot capture chip-to-chip communication effects.
In contrast, architecture-aware simulators like LLMServingSim~\cite{LLMServingSim}, Calculon~\cite{Calculon}, and LLMCompass~\cite{llmcompass} have made important progress in modeling computation efficiency and accelerator design, yet their communication modeling generally relies on analytical or simplified abstractions.

On the other hand, traditional NoC simulators such as BookSim~\cite{booksim} and Garnet~\cite{garnet3} (in gem5~\cite{gem5}) focus on on-chip interconnects, while CN-Sim~\cite{cnsim} and HexaMesh~\cite{Iff2023HexaMesh} target chiplet-level networks.
For system-level or scale-out networks, tools such as ns-3~\cite{ns3sigcomm}, OMNeT++~\cite{omnet} are commonly used. However, none of these can capture the hardware-level communication behavior of scaled-out systems with scaled-up chips~\cite{Graening2023Chiplets,Feng2022ChipletActuary}.

This work bridges gaps in prior studies by focusing on C2C communication within specific \text{LLM} workloads. It advances the field through a cycle-level, hardware-aware communication model, enabling detailed analysis of chip-to-chip interactions during both training and inference.

\begin{figure}[t]
	\centering
	\includegraphics[width=\linewidth]{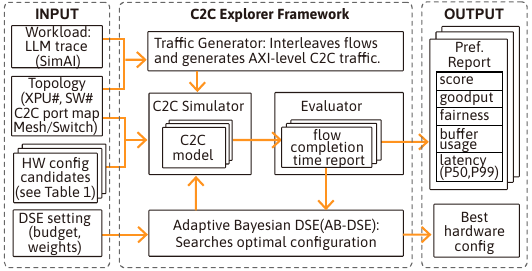}
	\caption{{\codename} framework workflow.}
	\label{fig:framework}
\end{figure}

\section{The \codename{} Framework}
Figure~\ref{fig:framework} depicts the proposed {\codename} framework.
It consists of 4 modules (from top to bottom): a \textit{traffic generator},
a \textit{C2C simulator}, an \textit{evaluator}, and an \textit{adaptive Bayesian
search engine} (\textit{AB-DSE}). Inputs for \codename{} include:
(i) LLM executing traces generated by SimAI~\cite{simai},
(ii) topology descriptions (XPU/switch (SW) counts, C2C port map, mesh or switched fabric),
(iii) candidate hardware parameters,
and (iv) exploration budget and metric weights.
Outputs of \codename{} are reports for system performance, including score, goodput, fairness, buffer, P50/P99 latency, and the combination of design parameters to achieve optimal performance.

The operating principle of \codename{} is: the \textit{traffic generator} emits AXI-level C2C traffic $\rightarrow$
the \textit{C2C simulator} evaluates each configuration $\rightarrow$ the \textit{evaluator} extracts flow-level metrics $\rightarrow$
\textit{AB-DSE} updates the search distribution. Hardware constraints (e.g., \texttt{chunk\_size}$\ge 2\cdot$~\texttt{MAC\_frame}) are applied before each step to prune infeasible designs.
This loop iterates until the preset iteration budget is reached.
The function and organization of each component are detailed as follows.

\subsection{Traffic Generator}\label{sec:traffic-generator}
\begin{figure}[!t]
\centering
\includegraphics[width=\linewidth]{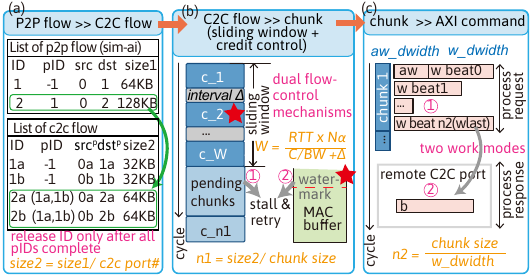}
\caption{
\textit{Traffic generator} workflow with dual flow control.
}
\label{fig:traffic-generator}
\end{figure}

C2C communication timing is governed by two interacting layers: (1) application-level message generation captured by SimAI, and (2) SoC-side burst pacing combined with link-level credit flow control. The \textit{traffic generator} embeds \texttt{sliding window} $+$ \texttt{credit control} to faithfully reproduce these cross-layer interactions.
As shown in Figure~\ref{fig:traffic-generator}, it transforms high-level LLM P2P traces into AXI-accurate C2C traffic through three tightly coupled stages: (a) P2P$\rightarrow$C2C flow mapping, (b) C2C$\rightarrow$chunk conversion with dual flow-control semantics, and(c) chunk$\rightarrow$AXI command emission.

\textbf{(a) P2P$\rightarrow$C2C flow mapping.}
SimAI~\cite{simai} provides XPU-level P2P traces containing message sizes and parent--child dependencies. Given the target C2C topology (full-mesh or switched fabric) and the number of C2C ports, the generator distributes each P2P flow across available ports. This produces multiple per-port sub-flows while preserving dependency order.

\textbf{(b) C2C$\rightarrow$chunk conversion with dual flow-control semantics.} Each C2C flow is segmented into fixed-size chunks (\(c\_1,c\_2,\dots\)), which are
injected into the pipeline under a \texttt{BDP-driven sliding window}. A new chunk is admitted only when (\texttt{active\_chunks} < W), where the window size is
\[
W=\Big\lceil \frac{BW \cdot RTT}{C}\, N \alpha \Big\rceil ,
\qquad
W=\Big\lceil \frac{RTT}{\tfrac{C}{BW}+\Delta}\, N \alpha \Big\rceil .
\]
\noindent
\textbf{Symbols:}
$BW$ (bandwidth), $RTT$ (round-trip time), $C$ (chunk size), $\Delta$ (inter-chunk gap), $N$ (parallel flows), $\alpha$ (safety factor).
Simultaneously, the physical-link behavior is modeled using \texttt{credit-based
backpressure}: when downstream MAC buffers hit their watermark, credits are
exhausted and the \textit{generator} stalls and retries every cycle.


\textbf{(c) Chunk$\rightarrow$AXI command emission.}
The \textit{generator} operates in two AXI-driven  modes. On the \textit{sender side}, each chunk is expanded into an AXI write burst: one \texttt{aw} request followed by $n_2=C/w_{d\_width}$ \texttt{w} beats, with the
final beat asserting \texttt{wlast}.
On the \textit{receiver side}, the remote C2C port detects \texttt{wlast} and
emits a \texttt{b} response to acknowledge burst completion. Receiving the
\texttt{b} response allows the sender to retire the chunk and decrement
\texttt{active\_chunks}.
This dual-mode behavior reproduces true AXI burst semantics and chunk-level
timing fidelity.

\subsection{C2C Simulator and C2C Model}\label{sec:c2c-model}
\textit{C2C simulator} covers both hardware modules (virtual channels, scheduling, flow control) and network behavior (Ethernet framing and switching), combining cycle-level port modeling with event-driven Ethernet/link modeling to balance timing fidelity and scalability. It is built on SimPy and accelerated with PyPy JIT~\cite{pypy1}.

\begin{figure*}[htbp]
\centering
\includegraphics[width=\linewidth]{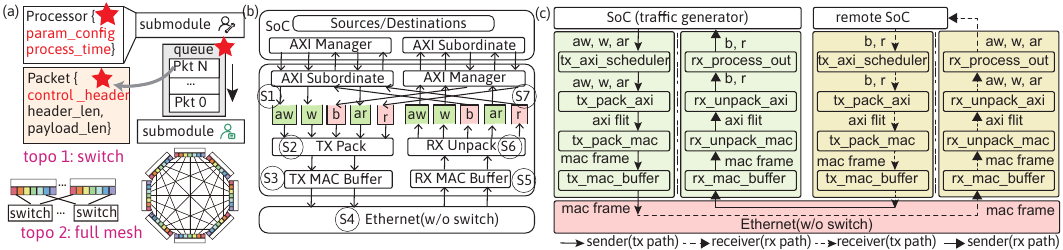}
\caption{(a) \textit{C2C simulator} features. (b) Target hardware architecture used in \textit{C2C simulator}. (c) Software pipeline loop. Each C2C port runs two concurrent tasks (TX and RX), while the Ethernet/switch is modeled as a shared global task, forming the end-to-end transfer loop.}
\label{fig:hardware}
\end{figure*}

Figure~\ref{fig:hardware}(a) summarizes the simulator’s three core features:

\begin{enumerate}[leftmargin=*,label=(\alph*)]
\item \textbf{Cycle-accurate execution in C2C port}: Each submodule is characterized by two parameters: param\_config (functional settings such as VC count or scheduling policy) and process\_time (cycles per operation).

\item \textbf{Packet-centric behavior characterization:}
The \textit{C2C simulator} adopts a packet-centric abstraction that matches the natural granularity of C2C communication. Each packet only keeps the fields that affect MAC/link behavior (\texttt{control\_header} plus \texttt{header\_len} and \texttt{payload\_len}), which suffice for transmission and framing calculations while omitting unnecessary intra-SoC details.

\item \textbf{Producer–consumer–style pipeline via FIFO decoupling}:
Submodules exchange packets through FIFO-decoupled queues, following a producer–consumer–style pipeline. Each stage consumes packets from its upstream queue and pushes results to its downstream FIFO every cycle, enabling a timing-synchronized multi-stage pipeline.

\end{enumerate}

Figure~\ref{fig:hardware}(b) shows the hardware architecture; Figure~\ref{fig:hardware}(c) presents its software-mapped counterpart in our simulator. In software, \textbf{each C2C port is instantiated as two concurrent tasks}: a \textit{TX task} executing the tx path (S1$\rightarrow$S2$\rightarrow$S3) and an \textit{RX task} executing the rx path (S5$\rightarrow$S6$\rightarrow$S7).
The Ethernet (w/o switch) is modeled as a separate shared task corresponding to stage S4, with which all ports interact during transmission.
Together, they form the full end-to-end transfer loop between C2C ports.

\begin{table}[!t]
\centering
\footnotesize
\caption{Parameters in Chip-to-Chip Model Submodules}
\label{tab:c2c_submodels_params}

\renewcommand{\arraystretch}{1.1}
\setlength{\tabcolsep}{3pt}

\begin{tabularx}{\columnwidth}{@{} c l l X @{}}
\specialrule{1pt}{0pt}{0pt}
Attr. & Submodule & Parameter & Value in Cases (Sec.~\ref{sec:usercase}) \\
\specialrule{1pt}{0pt}{0pt}

\threecir{\circred}{\circgray}{\circgray}
  & \texttt{traffic\_generator} & chunk\_size & 2KB, 4KB, 8KB \\

\threecir{\circgray}{\circgreen}{\circgray}
  & \texttt{tx\_axi\_scheduler} & AXI scheduler &
    DRR, LQ, RR, SP \\

\threecir{\circred}{\circgray}{\circgray}
  & \texttt{tx\_pack\_mac} & MAC\_frame\_size & 2KB, 4KB \\

\threecir{\circgray}{\circgreen}{\circblue}
  & \texttt{tx\_mac\_buffer} & VC\textunderscore pair\textunderscore nums & 1,2,4,8,16,32 \\

\threecir{\circgray}{\circgreen}{\circblue}
  & \texttt{tx\_mac\_buffer} & VC scheduler &
    DQD, FCFS, RR, WRR \\

\threecir{\circgray}{\circgray}{\circblue}
  & \texttt{rx\_mac\_buffer} & credit & 4KB, 8KB, 16KB, 32KB \\
\specialrule{1pt}{0pt}{0pt}
\multicolumn{4}{@{}p{\columnwidth}@{}}{{\footnotesize
Legend: \protect\circred\ Packetization \hfill
        \protect\circgreen\ Scheduling \hfill
        \protect\circblue\ Resource-Allocation.\hfill}}\\
\multicolumn{4}{@{}p{\columnwidth}@{}}{{\footnotesize
\textit{Notation:}
DRR = deficit round-robin (4\,KB quantum);
LQ = longest-queue-first;
RR = round-robin;
SP = strict priority (DA 0$\rightarrow$31);
DQD = DA-queue-depth;
FCFS = first-come first-served;
WRR = weighted round-robin (uniform).}}\\
\multicolumn{4}{@{}p{\columnwidth}@{}}{{\footnotesize
Our simulator exposes many tunable C2C parameters; Table~\ref{tab:c2c_submodels_params} lists the core ones used in this study.}}\\
\end{tabularx}
\end{table}

The simulator follows this path through a 7-stage pipeline (S1--S7) across multiple C2C ports, with each stage aligned to one or more evaluation axes--\textit{Packetization} (S2, S6), \textit{Scheduling} (S1, S3), and \textit{Resource-Allocation} (S3--S5). These mappings correspond directly to the attribute annotations in Table~\ref{tab:c2c_submodels_params}.

\textbf{(S1) SoC–C2C boundary scheduling [Scheduling]} .
AXI bursts triggered by the \textit{traffic generator} entering per-destination (DA)
queues. \texttt{tx\_axi\_scheduler} selects which DA to serve using configurable policies.
\textbf{(S2) TX Pack [Packetization].}
AXI channels (\texttt{aw, w, ar, r, b} in Figure~\ref{fig:hardware}(b)\&(c)) are packed into AXI flits in \texttt{tx\_pack\_axi}, and aggregated into MAC frames in \texttt{tx\_pack\_mac}.
Frames close upon DA changes to avoid cross-destination interleaving.
Request (\texttt{aw, w, r}) and response (\texttt{b, ar}) data paths
are physically separated to avoid bidirectional head-of-line blocking, especially under all-reduce traffic.
\textbf{(S3): TX MAC Buffer [Scheduling + Resource-Allocation].}
MAC frames are placed into per-VC buffers. In \texttt{tx\_mac\_buffer}, VC number, VC assignment, and VC scheduling determine the egress order. Credit-based flow control (CBFC)  throttles transmission when downstream buffers are full.
\textbf{(S4) Ethernet (w/o switch) [Resource-Allocation].}
For P2P links, this stage has fixed propagation delays plus per-frame serialization. When configured as a switch, it models an input-queued crossbar with per-(src,dst,vc) virtual output queues (VOQs) and round-robin arbitration. Cut-through is approximated via overlapping switch processing and link transmission. This stage captures VC utilization limits, credit availability, and multi-port contention.
\textbf{(S5) RX MAC Buffer [Resource-Allocation].}
Incoming MAC frames enter per-VC RX buffers. Upon admission, credits are
returned to the sender, completing the CBFC loop.
\textbf{(S6) RX Unpack [Packetization].}
MAC frames are decomposed into AXI flits and then into individual AXI
transactions. This reverses S2 and reconstitutes packet-level and
transaction-level structure on the receiver SoC.
\textbf{(S7) RX process output and completion signaling.}
The receiver issues AXI responses (\texttt{b}/\texttt{r}) to the remote sender.
A chunk completes only when the sender receives the final response, enabling accurate end-to-end timing measurement.

\subsection{Evaluator}\label{evaluator}
The \textit{C2C simulator} records each flow completion time (FCT), enabling the \textit{evaluator} to derive system-level communication metrics.
We assess each configuration from 5 dimensions: throughput, P50 latency, P99 latency, fairness, and buffer usage (Table~\ref{tab:evaluator}). Here, the buffer usage $N_V \cdot C_B$ reflects relative buffer provisioning, not literal hardware size. To obtain a scalar score $S$ that serves as the optimization objective for AB-DSE search algorithm, each metric is normalized and combined through a weighted sum, where goodput is directly rewarded and all other metrics are minimized. For the user case in Sec.~\ref{sec:usercase}, the weights prioritize throughput (40\%) and tail latency (25\%), reflecting the requirements of large-scale inference workloads.

\begin{table}[!b]
\caption{Evaluation metrics in {\codename}}
\label{tab:evaluator}
\centering
\footnotesize
\setlength{\tabcolsep}{3pt}
\begin{tabularx}{\columnwidth}{@{} l c c c c >{\raggedright\arraybackslash}X @{}}
\specialrule{1pt}{0pt}{0pt}
\textbf{Metric} & \textbf{Formula} & \textbf{Unit} & \textbf{Goal} & $w_m$ & \textbf{Description} \\
\specialrule{1pt}{0pt}{0pt}
Goodput       & ${\sum B_i }/{T_{\max} }$ & GBps   & $\uparrow$ & 0.40 & System throughput \\[4pt]
P50 Lat.           & $Q_{0.5}(\{F_i\})$            & cycles & $\downarrow$ & 0.15 & Median FCT \\[4pt]
P99 Lat.          & $Q_{0.99}(\{F_i\})$           & cycles & $\downarrow$ & 0.25 & 99th-percentile FCT \\[4pt]
Fairness  & $\sigma_F / \mu_F$            & --     & $\downarrow$ & 0.05 & Coeff. of variation \\[4pt]
Buffer      & $N_V \cdot C_B $              & KB     & $\downarrow$ & 0.15 & Buffer per port \\
\midrule
\multicolumn{6}{@{}p{\columnwidth}@{}}{\textbf{Score:} $S = \sum_m w_m \cdot \phi_m(\text{Norm}(m))$, where $\phi_{\text{Goodput}}(x)=x$, others $1-x$} \\
\specialrule{1pt}{0pt}{-1pt}
\multicolumn{6}{@{}p{\columnwidth}@{}}{\footnotesize{\textit{Notation:} $B_i$ = data size of flow $i$ (bytes); $T_{\max}$: max flow completion time (FCT, cycles); $F_i$: FCT of flow $i$; $Q_p$: $p$-quantile; $\sigma_F$, $\mu_F$: std. dev. and mean of all flows FCT; $N_V$: \#VCs; $C_B$: credit size; Norm$(m) = (m - m_{\min})/(m_{\max} - m_{\min})$.}}\\
\end{tabularx}
\end{table}


\subsection{Adaptive Bayesian Design Space Exploration}
With the score function defined in Section~\ref{evaluator}, the final step is to
search the large C2C design space for optimal combinations.
We develop an AB-DSE engine that tightly integrates an end-to-end workflow: couples hardware-aware constraints, cycle-accurate simulation, and metric-driven optimization.

As described in Algorithm~\ref{alg:ab_dse}, AB-DSE first applies
\textit{C2C-feasibility pruning} eliminates analytically dominated combinations
(e.g., suboptimal chunk--frame ratios), restricting the search to a more
efficient region of the design space~\cite{ZhangLYYZ0H19}~\cite{HuLH19}.
It then performs a diversity-oriented initialization using Latin Hypercube Sampling (LHS), where each sampled configuration is simulated in the \textit{C2C simulator} and scored in the \textit{evaluator}.
After this seeding phase, AB-DSE enters a Bayesian refinement loop. A Gaussian-process surrogate is trained on the accumulated history, and Expected Improvement (EI) selects the next candidate expected to improve goodput, latency, fairness, or buffer usage. The surrogate is updated iteratively.
When the exploration budget is exhausted, AB-DSE outputs the highest-scoring
configuration, representing the best achievable balance across all three C2C
axes for the given LLM traffic and topology~\cite{Lyu0YZ018}~\cite{ZhaiYWZ18}.

\begin{algorithm}[!h]
\caption{AB-DSE: C2C-Aware Adaptive Bayesian Search}
\label{alg:ab_dse}
\small
\begin{algorithmic}[1]
\Require Design space $\mathcal{X}$, budget $B$, ratio $\alpha$, traffic $T$
\Ensure Best configuration $x^\star$

\State $\mathcal{X}\!\gets\!\{x\!\in\!\mathcal{X}\mid\texttt{hw\_constraints}(x)\}$,
       $\mathcal{H}\!\gets\!\emptyset$,
       $B_{\text{init}}\!\gets\!\lfloor\alpha B\rfloor$

\Comment{\textbf{Init: LHS diversification}}
\For{$i=1$ to $B_{\text{init}}$}
    \State $x_i\!\sim\!\text{LHS}(\mathcal{X})$;
           $\mathbf{m}_i\!\gets\!\texttt{SimulateC2C}(x_i,T)$;
           $S_i\!\gets\!\texttt{Evaluate}(\mathbf{m}_i)$
    \State $\mathcal{H}\!\gets\!\mathcal{H}\cup\{(x_i,S_i)\}$
\EndFor

\Comment{\textbf{BO: C2C-guided refinement}}
\For{$i=B_{\text{init}}+1$ to $B$}
    \State Fit GP surrogate $g$ on $\mathcal{H}$
    \State $x_i\!\gets\!\arg\max_{x\in\mathcal{X}}\text{EI}(x;g,\mathcal{H})$
    \State $\mathbf{m}_i\!\gets\!\texttt{SimulateC2C}(x_i,T)$;
           $S_i\!\gets\!\texttt{Evaluate}(\mathbf{m}_i)$
    \State $\mathcal{H}\!\gets\!\mathcal{H}\cup\{(x_i,S_i)\}$
\EndFor

\State \Return $x^\star \gets \arg\max_{(x,S)\in\mathcal{H}} S$
\end{algorithmic}
\end{algorithm}

\section{Experimental Validation For C2C Simulator}
The proposed \textit{C2C simulator} achieves high emulation accuracy with reduced simulation time. The validation experiments and results are detailed as follows.

\subsection{Accuracy Evaluation}
\begin{figure}[!t]
\centering
\includegraphics[width=\linewidth]{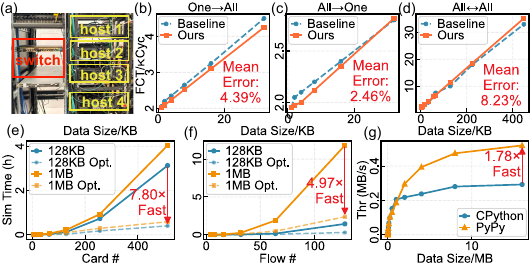}
\caption{(a) Baseline system (1$\times$ 400Gbps switch and 4$\times$ FPGA-based C2C host prototypes)~\cite{odcc_ethx_whitepaper_2025,odcc_ethx_testreport_2025} to verify \textit{C2C simulator}.
(b–d) Accuracy evaluation.
(e–f) Simulation time vs number of XPUs and number of flows for pure cycle model and hybrid models (optimization).
(g) Point-to-point simulation throughput for CPython and PyPy JIT~\cite{pypy1,pypy2}.}
\label{fig:exp}
\end{figure}
To the best of our knowledge, no existing simulator supports scale-up communication modeling. Following the ETH-X white-paper report, we configure our simulator using the same bandwidth link (400Gbps), switch forwarding delay (467 ns), and port processing latency (245 ns) to model the data link described in public technical reports~\cite{odcc_ethx_testreport_2025}.
To validate accuracy, we further built a real 400 Gbps system prototype consisting of 1 switch and 4 FPGA-based hosts (Figure~\ref{fig:exp}(a)). This platform exposes the same datapath behavior as ETH-X~\cite{odcc_ethx_whitepaper_2025}.
Figure~\ref{fig:exp}(b-d) evaluates a 4-XPU system under three representative patterns: (One$\rightarrow$All), (All$\rightarrow$One), and (All$\leftrightarrow$All).
The x-axis represents the message size (KB) being transmitted, and the y-axis represents the all flow completion time (cycle). Each scenario is swept across a range of message sizes (1\,KB--32\,KB for (b) and (c), and 1\,KB--448\,KB for (d)), resulting in multiple evaluation points per communication pattern. Across these scenarios, the mean error of all swept points relative to the ETH-X measurements is tolerably 4.39\%, 2.46\%, and 8.23\%, respectively.

\subsection{Performance and Scalability}
Hybrid modeling (combining \textsl{switch-event} scheduling with \textsl{C2C-port cycle} scheduling) in \codename{} can expedite large-scale simulation.
Figures~\ref{fig:exp}(e)–(f) evaluate this against a purely cycle-accurate baseline both on 2$\times$Intel Xeon Platinum 8358 (1TB main memory).
In Figure~\ref{fig:exp}(e), the x-axis is the number of concurrently communicating XPUs, while Figure~\ref{fig:exp}(f) uses the number of serialized flows.
Both report end-to-end simulation time on the y-axis.
For 128\,KB messages in Figure~\ref{fig:exp}(e), the hybrid scheduler achieves
1.1$\times$--7.8$\times$ speedups as parallelism scales from 4 to 512 XPUs (All$\leftrightarrow$All communication).
For 1\,MB messages, the corresponding speedups range from 0.98$\times$--6.9$\times$.
In Figure~\ref{fig:exp}(f), for 128\,KB messages across 4--128 serialized flows,
the hybrid scheduler achieves 0.77$\times$--4.8$\times$ speedups. For 1\,MB messages, the range is 0.72$\times$--5.0$\times$.
Figure~\ref{fig:exp}(g) reports P2P performance as message size increases on a 13th Gen Intel Core i7-1355U CPU (16GB main memory); the y-axis shows delivered data per unit simulation time. The PyPy JIT runtime provides an additional acceleration up to $1.78\times$@data size$\le$16\,MB. We expect the speedup ratio to increase as data sizes grow beyond 16MB based on the trend in Figure~\ref{fig:exp}(g).

\section{C2C Interconnect Design Space Exploration}
Based on the verified \textit{C2C simulator}, we use \codename{} to analyze the impact of \textit{Packetization}, \textit{Scheduling}, and \textit{Resource-Allocation} on C2C interconnect performance under LLM workloads.


\begin{figure}[!t]
\centering
\includegraphics[width=\linewidth]{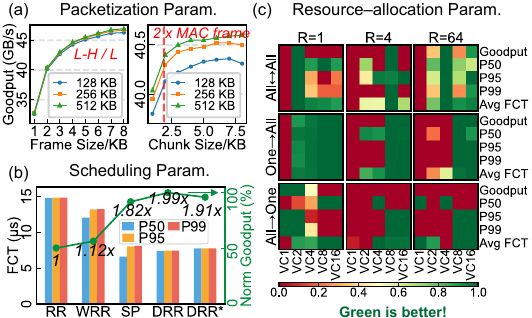}
\caption{
(a) Packetization parameters: effect of MAC frame size and chunk size on goodput.
(b) Comparison of AXI transmit–side schedulers: DRR/DRR* (deficit round-robin with 2\,KB/8\,KB quantum).
(c) Resource–allocation parameters: VC-pair count under three traffic imbalance levels
(R=1, 4, 64), where $R=\text{max}/\text{min}$ message size. Green indicates a higher normalized score.
}
\label{fig:exp-precision}
\end{figure}

\begin{figure*}[!t]
\centering
\includegraphics[width=\linewidth]{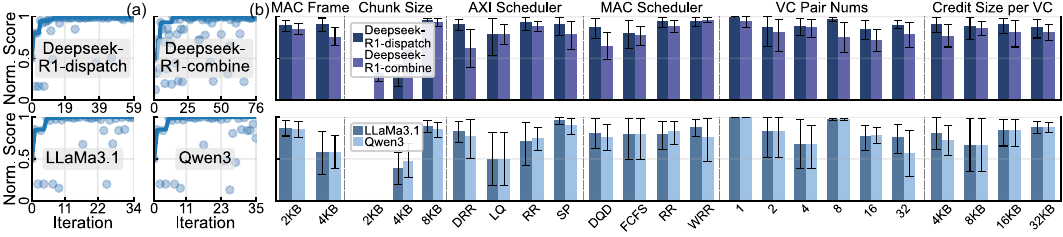}
\caption{User case, we evaluate \codename{} using four representative flow groups, including the \texttt{all-to-all} expert exchange in DeepSeek-R1-671B inference (\textit{dispatch} and \textit{combine}), the \texttt{all-reduce} synchronization in LLaMA3.1-405B inference, and the gradient \texttt{all-reduce} in Qwen3-30B training.
(Left) Normalized score vs iteration for four tasks.
(Right) Sensitivity of six design parameters with mean norm.\ scores and 95\% confidence intervals.}
\label{fig:exp-user}
\end{figure*}

\subsection{Packetization Parameters}\label{sec:packet}
In Figure~\ref{fig:exp-precision}(a), the left sub-figure x-axis sweeps the MAC frame size(KB), the right sub-figure x-axis varies the chunk size (with MAC frame size fixed at 1\,KB). The y-axis both reports the measured goodput (GB/s) under P2P (no-switch) transmission. The three curves correspond to total message sizes of 128\,KB, 256\,KB, and 512\,KB.
The goodput increases sharply as the MAC frame grows from 1\,KB to 8\,KB and then saturates, following the MAC efficiency $(L-H)/L$, where \(L\) is the MAC-frame length and \(H\) is the fixed per-frame protocol overhead. Increasing \(L\) amortizes \(H\), improving effective throughput.
Goodput improves with chunk size and plateaus once the chunk length reaches roughly twice the MAC-frame size. Each chunk triggers a standalone AXI write burst with its own address, control, and response phases; for small chunks, this per-transaction overhead dominates and depresses throughput. As chunk size increases, the overhead is amortized across more payload; goodput approaches saturation once both the AXI pipeline and the link are fully utilized.

\subsection{Scheduling Parameters}\label{sec:scheduler}
Figure~\ref{fig:exp-precision}(b) compares five AXI transmit–side schedulers in an
8-XPU (All$\leftrightarrow$All) experiment with 64\,KB messages. The x-axis lists the scheduling policies, and the y-axis reports latency (P50/P95/P99 FCT; lower is better) and normalized goodput (higher is better), enabling a joint comparison of throughput and latency under different scheduling strategies.
We evaluate: RR (equal round-robin), WRR (weighted RR; per-destination-address (DA) weights 8:4:2:1:1:1:1:1), SP (strict priority over DAs; priority order 0$\rightarrow$7), and DRR (deficit round-robin).
In DRR, the \emph{quantum} (2\,KB or 8\,KB) sets the maximum bytes each DA can transmit per service round; smaller quanta enforce fairness, while larger quanta permit longer bursts at the cost of higher HoL-blocking sensitivity. Relative to RR, both WRR and SP improve goodput (1.12$\times$ and 1.82$\times$), while DRR–2KB achieves the highest gain (1.99$\times$).
RR offers low unfairness but performs poorly because its frequent destination switching prevents MAC-frame aggregation, increasing per-frame overhead. In contrast, DRR transmits multiple packets to the same destination within one quantum, enabling larger MAC frames and higher link efficiency. However, enlarging the quantum from 2\,KB to 8\,KB slightly hurts performance (DRR–8KB reduces goodput by 4.06\%), as coarse-grained service lets one destination dominate a scheduling window and exacerbates head-of-line blocking (HoL blocking).

\begin{table}[b]
\centering
\caption{Performance improvements of the optimal configurations from AB-DSE over the worst feasible designs.}
\label{tab:opt_worst_5metrics}
\footnotesize
\setlength{\tabcolsep}{3pt}
\begin{tabularx}{\columnwidth}{@{} >{\raggedright\arraybackslash}X c c c c c @{}}
\specialrule{1pt}{0pt}{0pt}
\textbf{Task} &
\textbf{Goodput} &
\textbf{P50 Lat.} &
\textbf{P99 Lat.} &
\textbf{Fairness$^1$} &
\textbf{Buffer$^2$} \\
\hline
\textbf{dispatch}  & +14.7\% & --12.6\% &  --12.6\% & + 0.001 & --75\% \\
\textbf{combine} & +44.1\% & --30.4\% & --30.4\% & - & --98.4\% \\
\textbf{LLaMA3.1} & +51.7\% & --68.7\% & --68.7\% & -0.087 & --75\% \\
\textbf{Qwen3} & +50.5\% & --64.3\% & --64.3\% & -0.034 & --96.9\% \\
\specialrule{1pt}{0pt}{0pt}
\multicolumn{6}{@{}p{\columnwidth}@{}}{\footnotesize{\textit{Notation:}
\textit{Metric definitions are summarized in Table~\ref{tab:evaluator}.}
$^1$ Lower is better.
$^2$ Buffer Usage.}}
\end{tabularx}

\end{table}

\begin{table}[t]
\centering
\footnotesize
\caption{Comparison with representative simulators.}
\label{tab:compare_tools}
\setlength{\tabcolsep}{2pt}
\begin{tabularx}{\columnwidth}{@{}>{\raggedright\arraybackslash\hsize=0.7\hsize}X|>{\centering\arraybackslash\hsize=1.05\hsize}X>{\centering\arraybackslash\hsize=1.05\hsize}X>{\centering\arraybackslash\hsize=1.05\hsize}X>{\centering\arraybackslash\hsize=1.05\hsize}X>{\centering\arraybackslash\hsize=1.05\hsize}X>{\centering\arraybackslash\hsize=1.05\hsize}X@{}}
\specialrule{1pt}{0pt}{0pt}
\textbf{Feat.} &
\textbf{This Work} &
\textbf{bookSim \cite{booksim}} & %
\textbf{garnet \cite{garnet3}} & %
\textbf{cnsim \cite{cnsim}} & %
\textbf{ns-3 \cite{ns3sigcomm}} &
\textbf{OMNeT++ \cite{omnet}} \\
\specialrule{1pt}{0pt}{0pt}
LLM$^1$
&  \circblue & \circred & \circred & \circred & \circblue & \circred \\

FCT$^2$
& \circblue & \circred & \circred & \circred & partial & partial \\

TME$^3$
& hybrid & cycle & cycle & cycle & event & event \\

AXI$^4$
& \circblue & \circred & \circred & \circred & \circred & \circred \\

MP$^5$
& \circblue & \circred & \circred & \circred & \circred & \circred \\


CBFC$^6$
& \circblue & \circblue & \circblue & \circblue & \circred & \circred \\

SW$^7$
& \circblue & partial & partial & partial & \circblue & \circblue \\

\specialrule{1pt}{0pt}{0pt}
\multicolumn{7}{@{}p{\columnwidth}@{}}{\footnotesize
Legend: \circblue\ supported; \circred\ not supported.
$^{1}$LLM = support for real LLM communication traces.
$^{2}$FCT = support for flow completion time analysis.
$^{3}$TME = cycle/event/hybrid timing model.
$^{4}$AXI = support for AXI-level transaction modeling.
$^{5}$MP = MAC-layer packetization and dynamic frame construction; bookSim/garnet/cnsim use fixed flits only, while ns-3/OMNeT++ use pre-constructed MAC packets.
$^{6}$CBFC = credit-based flow control.
$^{7}$SW = switch-based topology.
} \\
\end{tabularx}
\end{table}

\subsection{Resource–Allocation Parameters}
Figure~\ref{fig:exp-precision}(c) quantifies how VC number impacts C2C performance across three communication patterns (All$\leftrightarrow$All, All$\rightarrow$One, One$\rightarrow$All) in an 8-XPU system.
Each heatmap cell shows the normalized score (green = higher), with VC pairs on the horizontal axis and evaluation metrics (Goodput, P50/P95/P99, Avg.~FCT) on the vertical axis.
We sweep three flow-size imbalance levels ($R=\text{max}/\text{min}$): balanced(64\,KB, $R=1$), light imbalance(32/64/128\,KB, $R=4$), and heavy imbalance (8/64/512\,KB, $R=64$).
The results reveal a clear trend: balanced traffic is nearly VC-insensitive; light imbalance benefits from 4--8 VC pairs; heavy imbalance collapses unless $N_V$ approaches the concurrent flow count.
This behavior follows from the effective imbalance per VC.
With global concurrency $C$ and $N_V$ VC pairs, each VC carries $\approx C/N_V$ flows, yielding $R_{\mathrm{eff}} \approx 1 + (R-1)\frac{(C/N_V)-1}{C-1}$.
As $N_V$ increases, $R_{\mathrm{eff}}\!\rightarrow\!1$, reducing HoL blocking and restoring throughput.
When $N_V$ is small, $R_{\mathrm{eff}}\!\approx\!R$, and imbalance dominates—especially in  All$\rightarrow$One, All$\leftrightarrow$All patterns.

\section{User Case Study}\label{sec:usercase}
We evaluate \codename{} on a 32-XPU system using three representative communication workloads that capture the dominant communication patterns in modern large-scale LLMs: the \texttt{all-to-all} expert exchange in DeepSeek-R1-671B inference~\cite{deepseek}, the \texttt{all-reduce} synchronization in LLaMA3.1-405B inference~\cite{llama3}, and the gradient \texttt{all-reduce} operations in Qwen3-30B training~\cite{qwen3}, with all flow traces derived from SimAI profiling.

The exploration covers 6 design parameters (Sec.~\ref{sec:c2c-model}), spanning packetization (MAC frame size, chunk size), scheduling (AXI-level scheduler, MAC-level scheduler), and resource-allocation (VC pair numbers, credit).
Figure~\ref{fig:exp-user}(a) shows the normalized score (y-axis) over exploration iterations (x-axis). The score weights follow Table~\ref{tab:evaluator}. Each blue dot corresponds to the score of an evaluated configuration, while the blue solid line tracks the best score achieved so far. All tasks converge within 20 evaluations. The packetization rule (\texttt{chunk\_size} $\ge 2\times$ \texttt{MAC\_frame}), used for all tasks except DeepSeek-R1 \textit{combine}, shrinks the feasible space from 2394 to 1152 configurations, removing low-quality points and accelerating convergence. The resulting optimal designs deliver the improvements summarized in Table~\ref{tab:opt_worst_5metrics}. Figure~\ref{fig:exp-user}(b) isolates per-parameter sensitivity. The x-axis enumerates that parameter’s candidate values, while the y-axis shows the mean normalized score with 95\% confidence intervals, and shorter bars reflect greater stability.
Parameter sensitivity reveals no monotonic trends, underscoring the need for joint optimization. Packetization is most influential, but larger frame size is not universally optimal across tasks. AXI-side scheduling has a greater impact than MAC-side arbitration. VC=1 often scores well due to minimal buffer cost, yet all workloads prefer multi-VC designs for balanced throughput and latency. Overall, optimal settings are strongly workload-dependent.



\section{Conclusion}

This work introduces \textit{C2C-Explorer}, a framework for hardware-aware design exploration of C2C interconnects in large-scale LLM workloads. It comprises a traffic generator, a C2C simulator, and an AB-DSE optimization engine, enabling simulation and optimization of configurations for multi-XPU interconnects. Experimental results validate \textit{C2C-Explorer}'s expedited execution speed and high accuracy (<9\%) and demonstrate its potential for exploring communication designs in scale-up and scale-out LLM computing systems.


\bibliographystyle{unsrtnat}
\bibliography{dac}

@article{liao2025ub,
  title={UB-Mesh: a Hierarchically Localized nD-FullMesh Datacenter Network Architecture},
  author={Liao, Heng and Liu, Bingyang and Chen, Xianping and Guo, Zhigang and Cheng, Chuanning and Wang, Jianbing and Chen, Xiangyu and Dong, Peng and Meng, Rui and Liu, Wenjie and others},
  journal={arXiv preprint arXiv:2503.20377},
  year={2025}
}

@misc{xu2025characterizingcommunicationpatternsdistributed,
      title={Characterizing Communication Patterns in Distributed Large Language Model Inference}, 
      author={Lang Xu and Kaushik Kandadi Suresh and Quentin Anthony and Nawras Alnaasan and Dhabaleswar K. Panda},
      year={2025},
      eprint={2507.14392},
      archivePrefix={arXiv},
      primaryClass={cs.DC},
      url={https://arxiv.org/abs/2507.14392}, 
}

@article{intto1,
author = {Cheng, Scott and Lin, Jun-Liang and Emani, Murali and Raskar, Siddhisanket and Foreman, Sam and Xie, Zhen and Vishwanath, Venkatram and Kandemir, Mahmut Taylan},
title = {Thorough Characterization and Analysis of Large Transformer Model Training At-Scale},
year = {2024},
issue_date = {March 2024},
publisher = {Association for Computing Machinery},
volume = {8},
number = {1},
url = {https://doi.org/10.1145/3639034},
doi = {10.1145/3639034},
month = feb,
articleno = {8},
numpages = {25}
}

@inproceedings {simai,
author = {Xizheng Wang and Qingxu Li and Yichi Xu and Gang Lu and Dan Li and Li Chen and Heyang Zhou and Linkang Zheng and Sen Zhang and Yikai Zhu and Yang Liu and Pengcheng Zhang and Kun Qian and Kunling He and Jiaqi Gao and Ennan Zhai and Dennis Cai and Binzhang Fu},
title = {{SimAI}: Unifying Architecture Design and Performance Tuning for {Large-Scale} Large Language Model Training with Scalability and Precision},
booktitle = {22nd USENIX Symposium on Networked Systems Design and Implementation, {NSDI} 2025)},
year = {2025},
isbn = {978-1-939133-46-5},
pages = {541--558},
url = {https://www.usenix.org/conference/nsdi25/presentation/wang-xizheng-simai},
publisher = {USENIX Association},
month = apr
}

@article{agrawal2024vidur,
  title={Vidur: A large-scale simulation framework for llm inference},
  author={Agrawal, Amey and Kedia, Nitin and Mohan, Jayashree and Panwar, Ashish and Kwatra, Nipun and Gulavani, Bhargav S and Ramjee, Ramachandran and Tumanov, Alexey},
  journal={In Proceedings of Machine Learning and Systems, 2024},
  volume={6},
  pages={351--366}
}

@inproceedings{LLMServingSim,
   title={LLMServingSim: A HW/SW Co-Simulation Infrastructure for LLM Inference Serving at Scale},
   url={http://dx.doi.org/10.1109/IISWC63097.2024.00012},
   DOI={10.1109/iiswc63097.2024.00012},
   booktitle={2024 IEEE International Symposium on Workload Characterization, {IISWC} 2024},
   publisher={IEEE},
   author={Cho, Jaehong and Kim, Minsu and Choi, Hyunmin and Heo, Guseul and Park, Jongse},
   year={2024},
   month=sep, pages={15–29} }

@inproceedings{Calculon,
author = {Isaev, Mikhail and Mcdonald, Nic and Dennison, Larry and Vuduc, Richard},
title = {Calculon: a methodology and tool for high-level co-design of systems and large language models},
year = {2023},
isbn = {9798400701092},
publisher = {Association for Computing Machinery},
url = {https://doi.org/10.1145/3581784.3607102},
booktitle = {Proceedings of the International Conference for High Performance Computing, Networking, Storage and Analysis},
articleno = {71},
numpages = {14},
series = {SC 2023}
}

@misc{llmcompass,
      title={A Hardware Evaluation Framework for Large Language Model Inference}, 
      author={Hengrui Zhang and August Ning and Rohan Prabhakar and David Wentzlaff},
      year={2023},
      eprint={2312.03134},
      archivePrefix={arXiv},
      primaryClass={cs.AR},
      url={https://arxiv.org/abs/2312.03134}, 
}

@inproceedings {cnsim,
author = {Yinxiao Feng and Yuchen Wei and Dong Xiang and Kaisheng Ma},
title = {Evaluating Chiplet-based {Large-Scale} Interconnection Networks via {Cycle-Accurate} {Packet-Parallel} Simulation},
booktitle = {2024 USENIX Annual Technical Conference, {USENIX} {ATC} 2024)},
year = {2024},
isbn = {978-1-939133-41-0},
pages = {731--747},
url = {https://www.usenix.org/conference/atc24/presentation/feng-yinxiao},
publisher = {USENIX Association},
month = jul
}

@INPROCEEDINGS{booksim,
  author={Nan Jiang and Becker, Daniel U. and Michelogiannakis, George and Balfour, James and Towles, Brian and Shaw, D. E. and Kim, John and Dally, William J.},
  booktitle={2013 IEEE International Symposium on Performance Analysis of Systems and Software, {ISPASS} 2013}, 
  title={A detailed and flexible cycle-accurate Network-on-Chip simulator}, 
  year={2013},
  volume={},
  number={},
  pages={86-96},
  doi={10.1109/ISPASS.2013.6557149},
  publisher={IEEE},
}

@INPROCEEDINGS{garnet3,
  author={Bharadwaj, Srikant and Yin, Jieming and Beckmann, Bradford and Krishna, Tushar},
  booktitle={2020 57th ACM/IEEE Design Automation Conference, {DAC} 2020}, 
  title={Kite: A Family of Heterogeneous Interposer Topologies Enabled via Accurate Interconnect Modeling}, 
  year={2020},
  volume={},
  number={},
  pages={1-6},
  publisher={ACM},
  doi={10.1109/DAC18072.2020.9218539}}

@inproceedings{ns3sigcomm,
  author    = {T. R. Henderson and M. Lacage and others},
  title     = {{ns-3}: A Discrete-Event Network Simulator for Internet Systems},
  booktitle = {Proceedings of the ACM SIGCOMM 2008 Conference on Data Communication},
  year      = {2008}
}

@inproceedings{AlawiehWL17,
  author       = {Mohamad Baker Alawieh  and 
                  Fa Wang  and 
                  Xin Li},
  title        ={{Efficient Hierarchical Performance Modeling for Integrated Circuits
                  via Bayesian Co-Learning}},
  booktitle    ={{Proceedings of the 54th Annual Design Automation Conference, {DAC}
                  2017}},
  pages        = {9:1--9:6},
  publisher    = {{ACM}},
  year         = {2017},
  doi          = {10.1145/3061639.3062235},
}

@inproceedings{WangYYZH17,
  author       = {Mengshuo Wang  and 
                  Fan Yang  and 
                  Changhao Yan  and 
                  Xuan Zeng  and 
                  Xiangdong Hu},
  title        ={{Efficient Bayesian Yield Optimization Approach for Analog  and  {SRAM}
                  Circuits}},
  booktitle    ={{Proceedings of the 54th Annual Design Automation Conference, {DAC}
                  2017}},
  pages        = {11:1--11:6},
  publisher    = {{ACM}},
  year         = {2017},
  doi          = {10.1145/3061639.3062234},
}

@inproceedings{MudassarKM18,
  author       = {Burhan Ahmad Mudassar  and 
                  Jong Hwan Ko  and 
                  Saibal Mukhopadhyay},
  title        = {{Edge-cloud collaborative processing for intelligent internet of things:
                  a case study on smart surveillance}},
  booktitle    = {Proceedings of the 55th Annual Design Automation Conference, {DAC}
                  2018},
  pages        = {146:1--146:6},
  publisher    = {{ACM}},
  year         = {2018},
  doi          = {10.1145/3195970.3196036},
}

@inproceedings{Lyu0YZ018,
  author       = {Wenlong Lyu  and 
                  Fan Yang  and 
                  Changhao Yan  and 
                  Dian Zhou  and 
                  Xuan Zeng},
  title        ={{Multi-objective Bayesian optimization for analog/RF circuit synthesis}},
  booktitle    ={{Proceedings of the 55th Annual Design Automation Conference, {DAC} 2018}},
  pages        = {11:1--11:6},
  publisher    = {{ACM}},
  year         = {2018},
  doi          = {10.1145/3195970.3196078}
}

@inproceedings{ZhaiYWZ18,
  author       = {Jinyuan Zhai  and 
                  Changhao Yan  and 
                  Sheng{-}Guo Wang  and 
                  Dian Zhou},
  title        ={{An efficient Bayesian yield estimation method for high dimensional
                   and  high sigma {SRAM} circuits}},
  booktitle    ={{Proceedings of the 55th Annual Design Automation Conference, {DAC}
                  2018}},
  pages        = {132:1--132:6},
  publisher    = {{ACM}},
  year         = {2018},
  doi          = {10.1145/3195970.3195987},
}

@inproceedings{HuLH19,
  author       = {Hanbin Hu  and 
                  Peng Li  and 
                  Jianhua Z. Huang},
  title        ={{Enabling High-Dimensional Bayesian Optimization for Efficient Failure
                  Detection of Analog  and  Mixed-Signal Circuits}},
  booktitle    ={{Proceedings of the 56th Annual Design Automation Conference,
                  {DAC} 2019}},
  pages        = {17},
  publisher    = {{ACM}},
  doi          = {10.1145/3316781.3317818}
}

@inproceedings{ZhangLYYZ0H19,
  author       = {Shuhan Zhang  and 
                  Wenlong Lyu  and 
                  Fan Yang  and 
                  Changhao Yan  and 
                  Dian Zhou  and 
                  Xuan Zeng  and 
                  Xiangdong Hu},
  title        = {{An Efficient Multi-fidelity Bayesian Optimization Approach for Analog
                  Circuit Synthesis}},
  booktitle    = {Proceedings of the 56th Annual Design Automation Conference 2019,
                  {DAC} 2019},
  pages        = {64},
  publisher    = {{ACM}},
  year         = {2019},
  doi          = {10.1145/3316781.3317765}
}

@inproceedings{MaoYLLC19,
  author       = {Jiachen Mao  and 
                  Qing Yang  and 
                  Ang Li  and 
                  Hai Helen Li  and 
                  Yiran Chen},
  title        ={{MobiEye: An Efficient Cloud-based Video Detection System for Real-time
                  Mobile Applications}},
  booktitle    ={{Proceedings of the 56th Annual Design Automation Conference 2019,
                  {DAC} 2019}},
  pages        = {102},
  publisher    = {{ACM}},
  year         = {2019},
  doi          = {10.1145/3316781.3317865},
}

@inproceedings{pypy1,
author = {Bolz, Carl Friedrich and Cuni, Antonio and Fijalkowski, Maciej and Rigo, Armin},
title = {Tracing the meta-level: PyPy's tracing JIT compiler},
year = {2009},
isbn = {9781605585413},
publisher = {Association for Computing Machinery},
url = {https://doi.org/10.1145/1565824.1565827},
doi = {10.1145/1565824.1565827},
pages = {18–25},
numpages = {8},
location = {Genova, Italy},
series = {ICOOOLPS 2009}
}

@article{pypy2,
author = {Ard\"{o}, H\r{a}kan and Bolz, Carl Friedrich and FijaBkowski, Maciej},
title = {Loop-aware optimizations in PyPy's tracing JIT},
year = {2012},
issue_date = {February 2013},
publisher = {Association for Computing Machinery},
volume = {48},
number = {2},
issn = {0362-1340},
doi = {10.1145/2480360.2384586},
month = oct,
pages = {63–72},
numpages = {10},
}

@article{deepseek,
  author       = {Daya Guo and Dejian Yang and Haowei Zhang and Junxiao Song and Peiyi Wang and Qihao Zhu and Runxin Xu and Ruoyu Zhang and Shirong Ma and Xiao Bi and Xiaokang Zhang and Xingkai Yu and Yu Wu and Z. F. Wu and Zhibin Gou and Zhihong Shao and Zhuoshu Li and Ziyi Gao and Aixin Liu and Bing Xue and Bingxuan Wang and Bochao Wu and Bei Feng and Chengda Lu and Chenggang Zhao and Chengqi Deng and Chong Ruan and Damai Dai and Deli Chen and Dongjie Ji and Erhang Li and Fangyun Lin and Fucong Dai and Fuli Luo and Guangbo Hao and Guanting Chen and Guowei Li and H. Zhang and Hanwei Xu and Honghui Ding and Huazuo Gao and Hui Qu and Hui Li and Jianzhong Guo and Jiashi Li and Jingchang Chen and Jingyang Yuan and Jinhao Tu and Junjie Qiu and Junlong Li and J. L. Cai and Jiaqi Ni and Jian Liang and Jin Chen and Kai Dong and Kai Hu and Kaichao You and Kaige Gao and Kang Guan and Kexin Huang and Kuai Yu and Lean Wang and Lecong Zhang and Liang Zhao and Litong Wang and Liyue Zhang and Lei Xu and Leyi Xia and Mingchuan Zhang and Minghua Zhang and Minghui Tang and Mingxu Zhou and Meng Li and Miaojun Wang and Mingming Li and Ning Tian and Panpan Huang and Peng Zhang and Qiancheng Wang and Qinyu Chen and Qiushi Du and Ruiqi Ge and Ruisong Zhang and Ruizhe Pan and Runji Wang and R. J. Chen and R. L. Jin and Ruyi Chen and Shanghao Lu and Shangyan Zhou and Shanhuang Chen and Shengfeng Ye and Shiyu Wang and Shuiping Yu and Shunfeng Zhou and Shuting Pan and S. S. Li and Shuang Zhou and Shaoqing Wu and Tao Yun and Tian Pei and Tianyu Sun and T. Wang and Wangding Zeng and Wen Liu and Wenfeng Liang and Wenjun Gao and Wenqin Yu and Wentao Zhang and W. L. Xiao and Wei An and Xiaodong Liu and Xiaohan Wang and Xiaokang Chen and Xiaotao Nie and Xin Cheng and Xin Liu and Xin Xie and Xingchao Liu and Xinyu Yang and Xinyuan Li and Xuecheng Su and Xuheng Lin and X. Q. Li and Xiangyue Jin and Xiaojin Shen and Xiaosha Chen and Xiaowen Sun and Xiaoxiang Wang and Xinnan Song and Xinyi Zhou and Xianzu Wang and Xinxia Shan and Y. K. Li and Y. Q. Wang and Y. X. Wei and Yang Zhang and Yanhong Xu and Yao Li and Yao Zhao and Yaofeng Sun and Yaohui Wang and Yi Yu and Yichao Zhang and Yifan Shi and Yiliang Xiong and Ying He and Yishi Piao and Yisong Wang and Yixuan Tan and Yiyang Ma and Yiyuan Liu and Yongqiang Guo and Yuan Ou and Yuduan Wang and Yue Gong and Yuheng Zou and Yujia He},
  title        = {DeepSeek‐R1 incentivizes reasoning in LLMs through reinforcement learning},
  journal      = {Nature},
  volume       = {645},
  number       = {8081},
  pages        = {633--638},
  year         = {2025},
  doi          = {10.1038/s41586-025-09422-z}
}

@misc{megatron,
      title={Megatron-LM: Training Multi-Billion Parameter Language Models Using Model Parallelism}, 
      author={Mohammad Shoeybi and Mostofa Patwary and Raul Puri and Patrick LeGresley and Jared Casper and Bryan Catanzaro},
      year={2020},
      eprint={1909.08053},
      archivePrefix={arXiv},
      primaryClass={cs.CL},
      url={https://arxiv.org/abs/1909.08053}, 
}

@inproceedings{deepspeed,
author = {Rasley, Jeff and Rajbhandari, Samyam and Ruwase, Olatunji and He, Yuxiong},
title = {DeepSpeed: System Optimizations Enable Training Deep Learning Models with Over 100 Billion Parameters},
year = {2020},
isbn = {9781450379984},
publisher = {Association for Computing Machinery},
url = {https://doi.org/10.1145/3394486.3406703},
doi = {10.1145/3394486.3406703},
booktitle = {Proceedings of the 26th ACM SIGKDD International Conference on Knowledge Discovery \& Data Minin, KDD 2020},
pages = {3505–3506},
numpages = {2}
}

@misc{NCCL,
  author       = {NVIDIA Corporation},
  title        = {NCCL: NVIDIA Collective Communications Library},
  year         = {2025},
  howpublished = {\url{https://developer.nvidia.com/nccl}},
  note         = {Accessed: 2025-11-18},
}

@article{gem5,
author = {Binkert, Nathan and Beckmann, Bradford and Black, Gabriel and Reinhardt, Steven K. and Saidi, Ali and Basu, Arkaprava and Hestness, Joel and Hower, Derek R. and Krishna, Tushar and Sardashti, Somayeh and Sen, Rathijit and Sewell, Korey and Shoaib, Muhammad and Vaish, Nilay and Hill, Mark D. and Wood, David A.},
title = {The gem5 simulator},
year = {2011},
issue_date = {May 2011},
publisher = {Association for Computing Machinery},
volume = {39},
number = {2},
issn = {0163-5964},
url = {https://doi.org/10.1145/2024716.2024718},
doi = {10.1145/2024716.2024718},
journal = {SIGARCH Comput. Archit. News},
month = aug,
pages = {1–7},
numpages = {7}
}

@inproceedings{omnet,
author = {Varga, Andr\'{a}s and Hornig, Rudolf},
title = {An overview of the OMNeT++ simulation environment},
year = {2008},
isbn = {9789639799202},
publisher = {ICST (Institute for Computer Sciences, Social-Informatics and Telecommunications Engineering)},
booktitle = {Proceedings of the 1st International Conference on Simulation Tools and Techniques for Communications, Networks and Systems \& Workshops},
articleno = {60},
numpages = {10},
location = {Marseille, France},
series = {Simutools '08}
}

@techreport{odcc_ethx_whitepaper_2025,
  title        = {ETH-X Scale Up Interconnect Protocol White Paper, Version 1.0},
  author       = {{Open Data Center Committee (ODCC)}},
  institution  = {Open Data Center Committee (ODCC)},
  number       = {ODCC-2025-03002},
  year         = {2025},
  month        = sep,
  location     = {China},
  note         = {In Chinese}
}

@techreport{odcc_ethx_testreport_2025,
  title        = {ETH-X Scale Up Interconnect Protocol Test Report},
  author       = {{Open Data Center Committee (ODCC)}},
  institution  = {Open Data Center Committee (ODCC)},
  number       = {ODCC-2025-03005},
  year         = {2025},
  month        = sep,
  location     = {China},
  note         = {In Chinese}
}

@inproceedings{Iff2023HexaMesh,
  author    = {Patrick Iff and Maciej Besta and Matheus Cavalcante and Tim Fischer and Luca Benini and Torsten Hoefler},
  title     = {HexaMesh: Scaling to Hundreds of Chiplets with an Optimized Chiplet Arrangement},
  booktitle = {Proceedings of the 60th ACM/IEEE Design Automation Conference, {DAC} 2023},
  year      = {2023},
  pages     = {1--6},
  doi       = {10.1109/DAC56929.2023.10248006}
}

@inproceedings{Graening2023Chiplets,
  author    = {Alexander Graening and Saptadeep Pal and Puneet Gupta},
  title     = {{Chiplets: How Small is Too Small?}},
  booktitle = {Proceedings of the 60th ACM/IEEE Design Automation Conference, {DAC} 2023},
  year      = {2023},
  pages     = {1--6},
  doi       = {10.1109/DAC56929.2023.10247947}
}

@inproceedings{Kim2019CoDesign2.5D,
  author    = {Jinwoo Kim and Gauthaman Murali and Heechun Park and Eric Qin and Hyoukjun Kwon and Venkata Chaitanya and Krishna Chekuri and Nihar Dasari and Arvind Singh and Minah Lee and Hakki Mert Torun and Kallol Roy and Madhavan Swaminathan and Saibal Mukhopadhyay and Tushar Krishna and Sung Kyu Lim},
  title     = {{Architecture, Chip, and Package Co-design Flow for 2.5D IC Design Enabling Heterogeneous {IP} {R}euse}},
  booktitle = {Proceedings of the 56th Annual Design Automation Conference, {DAC} 2019},
  year      = {2019},
  pages     = {Article 178, 6 pages},
  doi       = {10.1145/3316781.3317775}
}

@inproceedings{Feng2022ChipletActuary,
author = {Yinxiao Feng and Kaisheng Ma},
title = {{Chiplet Actuary: A Quantitative Cost Model and Multi-Chiplet Architecture Exploration}},
booktitle = {Proceedings of the 59th ACM/IEEE Design Automation Conference, DAC 2022},
year = {2022},
pages = {121--126},
doi = {10.1145/3489517.3530428},
publisher = {ACM},
}

@inproceedings{vtrain,
  title={vtrain: A simulation framework for evaluating cost-effective and compute-optimal large language model training},
  author={Bang, Jehyeon and Choi, Yujeong and Kim, Myeongwoo and Kim, Yongdeok and Rhu, Minsoo},
  booktitle={2024 57th IEEE/ACM International Symposium on Microarchitecture, {MICRO} 2024},
  pages={153--167},
  year={2024},
  doi = {10.1109/MICRO61859.2024.00021},
  organization={IEEE}
}

@misc{llama3,
      title={The Llama 3 Herd of Models}, 
      author={Aaron Grattafiori and Abhimanyu Dubey and Abhinav Jauhri and Abhinav Pandey and Abhishek Kadian and Ahmad Al-Dahle and Aiesha Letman and Akhil Mathur and Alan Schelten and Alex Vaughan and Amy Yang and Angela Fan and Anirudh Goyal and Anthony Hartshorn and Aobo Yang and Archi Mitra and Archie Sravankumar and Artem Korenev and Arthur Hinsvark and Arun Rao and Aston Zhang and Aurelien Rodriguez and Austen Gregerson and Ava Spataru and Baptiste Roziere and Bethany Biron and Binh Tang and Bobbie Chern and Charlotte Caucheteux and Chaya Nayak and Chloe Bi and Chris Marra and Chris McConnell and Christian Keller and Christophe Touret and Chunyang Wu and Corinne Wong and Cristian Canton Ferrer and Cyrus Nikolaidis and Damien Allonsius and Daniel Song and Danielle Pintz and Danny Livshits and Danny Wyatt and David Esiobu and Dhruv Choudhary and Dhruv Mahajan and Diego Garcia-Olano and Diego Perino and Dieuwke Hupkes and Egor Lakomkin and Ehab AlBadawy and Elina Lobanova and Emily Dinan and Eric Michael Smith and Filip Radenovic and Francisco Guzmán and Frank Zhang and Gabriel Synnaeve and Gabrielle Lee and Georgia Lewis Anderson and Govind Thattai and Graeme Nail and Gregoire Mialon and Guan Pang and Guillem Cucurell and Hailey Nguyen and Hannah Korevaar and Hu Xu and Hugo Touvron and Iliyan Zarov and Imanol Arrieta Ibarra and Isabel Kloumann and Ishan Misra and Ivan Evtimov and Jack Zhang and Jade Copet and Jaewon Lee and Jan Geffert and Jana Vranes and Jason Park and Jay Mahadeokar and Jeet Shah and Jelmer van der Linde and Jennifer Billock and Jenny Hong and Jenya Lee and Jeremy Fu and Jianfeng Chi and Jianyu Huang and Jiawen Liu and Jie Wang and Jiecao Yu and Joanna Bitton and Joe Spisak and Jongsoo Park and Joseph Rocca and Joshua Johnstun and Joshua Saxe and Junteng Jia and Kalyan Vasuden Alwala and Karthik Prasad and Kartikeya Upasani and Kate Plawiak and Ke Li and Kenneth Heafield and Kevin Stone and Khalid El-Arini and Krithika Iyer and Kshitiz Malik and Kuenley Chiu and Kunal Bhalla and Kushal Lakhotia and Lauren Rantala-Yeary and Laurens van der Maaten and Lawrence Chen and Liang Tan and Liz Jenkins and Louis Martin and Lovish Madaan and Lubo Malo and Lukas Blecher and Lukas Landzaat and Luke de Oliveira and Madeline Muzzi and Mahesh Pasupuleti and Mannat Singh and Manohar Paluri and Marcin Kardas and Maria Tsimpoukelli and Mathew Oldham and Mathieu Rita and Maya Pavlova and Melanie Kambadur and Mike Lewis and Min Si and Mitesh Kumar Singh and Mona Hassan and Naman Goyal and Narjes Torabi and Nikolay Bashlykov and Nikolay Bogoychev and Niladri Chatterji and Ning Zhang and Olivier Duchenne and Onur Çelebi and Patrick Alrassy and Pengchuan Zhang and Pengwei Li and Petar Vasic and Peter Weng and Prajjwal Bhargava and Pratik Dubal and Praveen Krishnan and Punit Singh Koura and Puxin Xu and Qing He and Qingxiao Dong and Ragavan Srinivasan and Raj Ganapathy and Ramon Calderer and Ricardo Silveira Cabral and Robert Stojnic and Roberta Raileanu and Rohan Maheswari and Rohit Girdhar and Rohit Patel and Romain Sauvestre and Ronnie Polidoro and Roshan Sumbaly and Ross Taylor and Ruan Silva and Rui Hou and Rui Wang and Saghar Hosseini and Sahana Chennabasappa and Sanjay Singh and Sean Bell and Seohyun Sonia Kim and Sergey Edunov and Shaoliang Nie and Sharan Narang and Sharath Raparthy and Sheng Shen and Shengye Wan and Shruti Bhosale and Shun Zhang and Simon Vandenhende and Soumya Batra and Spencer Whitman and Sten Sootla and Stephane Collot and Suchin Gururangan and Sydney Borodinsky and Tamar Herman and Tara Fowler and Tarek Sheasha and Thomas Georgiou and Thomas Scialom and Tobias Speckbacher and Todor Mihaylov and Tong Xiao and Ujjwal Karn and Vedanuj Goswami and Vibhor Gupta and Vignesh Ramanathan and Viktor Kerkez and Vincent Gonguet and Virginie Do and Vish Vogeti and Vítor Albiero and Vladan Petrovic and Weiwei Chu and Wenhan Xiong and Wenyin Fu and Whitney Meers and Xavier Martinet and Xiaodong Wang and Xiaofang Wang and Xiaoqing Ellen Tan and Xide Xia and Xinfeng Xie and Xuchao Jia and Xuewei Wang and Yaelle Goldschlag and Yashesh Gaur and Yasmine Babaei and Yi Wen and Yiwen Song and Yuchen Zhang and Yue Li and Yuning Mao and Zacharie Delpierre Coudert and Zheng Yan and Zhengxing Chen and Zoe Papakipos and Aaditya Singh and Aayushi Srivastava and Abha Jain and Adam Kelsey and Adam Shajnfeld and Adithya Gangidi and Adolfo Victoria and Ahuva Goldstand and Ajay Menon and Ajay Sharma and Alex Boesenberg and Alexei Baevski and Allie Feinstein and Amanda Kallet and Amit Sangani and Amos Teo and Anam Yunus and Andrei Lupu and Andres Alvarado and Andrew Caples and Andrew Gu and Andrew Ho and Andrew Poulton and Andrew Ryan and Ankit Ramchandani and Annie Dong and Annie Franco and Anuj Goyal and Aparajita Saraf and Arkabandhu Chowdhury and Ashley Gabriel and Ashwin Bharambe and Assaf Eisenman and Azadeh Yazdan and Beau James and Ben Maurer and Benjamin Leonhardi and Bernie Huang and Beth Loyd and Beto De Paola and Bhargavi Paranjape and Bing Liu and Bo Wu and Boyu Ni and Braden Hancock and Bram Wasti and Brandon Spence and Brani Stojkovic and Brian Gamido and Britt Montalvo and Carl Parker and Carly Burton and Catalina Mejia and Ce Liu and Changhan Wang and Changkyu Kim and Chao Zhou and Chester Hu and Ching-Hsiang Chu and Chris Cai and Chris Tindal and Christoph Feichtenhofer and Cynthia Gao and Damon Civin and Dana Beaty and Daniel Kreymer and Daniel Li and David Adkins and David Xu and Davide Testuggine and Delia David and Devi Parikh and Diana Liskovich and Didem Foss and Dingkang Wang and Duc Le and Dustin Holland and Edward Dowling and Eissa Jamil and Elaine Montgomery and Eleonora Presani and Emily Hahn and Emily Wood and Eric-Tuan Le and Erik Brinkman and Esteban Arcaute and Evan Dunbar and Evan Smothers and Fei Sun and Felix Kreuk and Feng Tian and Filippos Kokkinos and Firat Ozgenel and Francesco Caggioni and Frank Kanayet and Frank Seide and Gabriela Medina Florez and Gabriella Schwarz and Gada Badeer and Georgia Swee and Gil Halpern and Grant Herman and Grigory Sizov and Guangyi and Zhang and Guna Lakshminarayanan and Hakan Inan and Hamid Shojanazeri and Han Zou and Hannah Wang and Hanwen Zha and Haroun Habeeb and Harrison Rudolph and Helen Suk and Henry Aspegren and Hunter Goldman and Hongyuan Zhan and Ibrahim Damlaj and Igor Molybog and Igor Tufanov and Ilias Leontiadis and Irina-Elena Veliche and Itai Gat and Jake Weissman and James Geboski and James Kohli and Janice Lam and Japhet Asher and Jean-Baptiste Gaya and Jeff Marcus and Jeff Tang and Jennifer Chan and Jenny Zhen and Jeremy Reizenstein and Jeremy Teboul and Jessica Zhong and Jian Jin and Jingyi Yang and Joe Cummings and Jon Carvill and Jon Shepard and Jonathan McPhie and Jonathan Torres and Josh Ginsburg and Junjie Wang and Kai Wu and Kam Hou U and Karan Saxena and Kartikay Khandelwal and Katayoun Zand and Kathy Matosich and Kaushik Veeraraghavan and Kelly Michelena and Keqian Li and Kiran Jagadeesh and Kun Huang and Kunal Chawla and Kyle Huang and Lailin Chen and Lakshya Garg and Lavender A and Leandro Silva and Lee Bell and Lei Zhang and Liangpeng Guo and Licheng Yu and Liron Moshkovich and Luca Wehrstedt and Madian Khabsa and Manav Avalani and Manish Bhatt and Martynas Mankus and Matan Hasson and Matthew Lennie and Matthias Reso and Maxim Groshev and Maxim Naumov and Maya Lathi and Meghan Keneally and Miao Liu and Michael L. Seltzer and Michal Valko and Michelle Restrepo and Mihir Patel and Mik Vyatskov and Mikayel Samvelyan and Mike Clark and Mike Macey and Mike Wang and Miquel Jubert Hermoso and Mo Metanat and Mohammad Rastegari and Munish Bansal and Nandhini Santhanam and Natascha Parks and Natasha White and Navyata Bawa and Nayan Singhal and Nick Egebo and Nicolas Usunier and Nikhil Mehta and Nikolay Pavlovich Laptev and Ning Dong and Norman Cheng and Oleg Chernoguz and Olivia Hart and Omkar Salpekar and Ozlem Kalinli and Parkin Kent and Parth Parekh and Paul Saab and Pavan Balaji and Pedro Rittner and Philip Bontrager and Pierre Roux and Piotr Dollar and Polina Zvyagina and Prashant Ratanchandani and Pritish Yuvraj and Qian Liang and Rachad Alao and Rachel Rodriguez and Rafi Ayub and Raghotham Murthy and Raghu Nayani and Rahul Mitra and Rangaprabhu Parthasarathy and Raymond Li and Rebekkah Hogan and Robin Battey and Rocky Wang and Russ Howes and Ruty Rinott and Sachin Mehta and Sachin Siby and Sai Jayesh Bondu and Samyak Datta and Sara Chugh and Sara Hunt and Sargun Dhillon and Sasha Sidorov and Satadru Pan and Saurabh Mahajan and Saurabh Verma and Seiji Yamamoto and Sharadh Ramaswamy and Shaun Lindsay and Shaun Lindsay and Sheng Feng and Shenghao Lin and Shengxin Cindy Zha and Shishir Patil and Shiva Shankar and Shuqiang Zhang and Shuqiang Zhang and Sinong Wang and Sneha Agarwal and Soji Sajuyigbe and Soumith Chintala and Stephanie Max and Stephen Chen and Steve Kehoe and Steve Satterfield and Sudarshan Govindaprasad and Sumit Gupta and Summer Deng and Sungmin Cho and Sunny Virk and Suraj Subramanian and Sy Choudhury and Sydney Goldman and Tal Remez and Tamar Glaser and Tamara Best and Thilo Koehler and Thomas Robinson and Tianhe Li and Tianjun Zhang and Tim Matthews and Timothy Chou and Tzook Shaked and Varun Vontimitta and Victoria Ajayi and Victoria Montanez and Vijai Mohan and Vinay Satish Kumar and Vishal Mangla and Vlad Ionescu and Vlad Poenaru and Vlad Tiberiu Mihailescu and Vladimir Ivanov and Wei Li and Wenchen Wang and Wenwen Jiang and Wes Bouaziz and Will Constable and Xiaocheng Tang and Xiaojian Wu and Xiaolan Wang and Xilun Wu and Xinbo Gao and Yaniv Kleinman and Yanjun Chen and Ye Hu and Ye Jia and Ye Qi and Yenda Li and Yilin Zhang and Ying Zhang and Yossi Adi and Youngjin Nam and Yu and Wang and Yu Zhao and Yuchen Hao and Yundi Qian and Yunlu Li and Yuzi He and Zach Rait and Zachary DeVito and Zef Rosnbrick and Zhaoduo Wen and Zhenyu Yang and Zhiwei Zhao and Zhiyu Ma},
      year={2024},
      eprint={2407.21783},
      archivePrefix={arXiv},
      primaryClass={cs.AI},
      url={https://arxiv.org/abs/2407.21783}, 
}

@misc{qwen3,
      title={Qwen3 Technical Report}, 
      author={An Yang and Anfeng Li and Baosong Yang and Beichen Zhang and Binyuan Hui and Bo Zheng and Bowen Yu and Chang Gao and Chengen Huang and Chenxu Lv and Chujie Zheng and Dayiheng Liu and Fan Zhou and Fei Huang and Feng Hu and Hao Ge and Haoran Wei and Huan Lin and Jialong Tang and Jian Yang and Jianhong Tu and Jianwei Zhang and Jianxin Yang and Jiaxi Yang and Jing Zhou and Jingren Zhou and Junyang Lin and Kai Dang and Keqin Bao and Kexin Yang and Le Yu and Lianghao Deng and Mei Li and Mingfeng Xue and Mingze Li and Pei Zhang and Peng Wang and Qin Zhu and Rui Men and Ruize Gao and Shixuan Liu and Shuang Luo and Tianhao Li and Tianyi Tang and Wenbiao Yin and Xingzhang Ren and Xinyu Wang and Xinyu Zhang and Xuancheng Ren and Yang Fan and Yang Su and Yichang Zhang and Yinger Zhang and Yu Wan and Yuqiong Liu and Zekun Wang and Zeyu Cui and Zhenru Zhang and Zhipeng Zhou and Zihan Qiu},
      year={2025},
      eprint={2505.09388},
      archivePrefix={arXiv},
      primaryClass={cs.CL},
      url={https://arxiv.org/abs/2505.09388}, 
}

\end{document}